\documentclass[useAMS,usenatbib]{mn2e}

\usepackage[latin1]{inputenc}
\usepackage{aas_macros}
\usepackage{txfonts}

\newcommand{\un}[1]{\mbox{ \rmfamily #1}}
\newcommand{\unp}[1]{\mbox{\rmfamily #1}}
\newcommand{\url}[1]{\ttfamily #1\normalfont}
\newcommand{\tab}[1]{Table~\ref{#1}}
\newcommand{\undeg}{\mbox{\textdegree}}
\newcommand{\unsec}{''}

\title[Relation between 2003~EH1, C/1490~Y1 and C/1385~U1]{
Updated analysis of the dynamical relation between asteroid 2003~EH1 and comets C/1490~Y1 and C/1385~U1
}

\author[M.~Micheli, F.~Bernardi, D.~J.~Tholen]
{
Marco Micheli$^1$, Fabrizio Bernardi$^1$, David J. Tholen$^1$\\
$^1$Institute for Astronomy, University of Hawaii, 2680 Woodlawn Drive, HI 96822 Honolulu, USA\\
}

\begin{document}

\maketitle

\begin{abstract}
The asteroid 2003~EH1, proposed as the parent body of the Quadrantid meteor shower, is thought to be the remnant of a past cometary object, tentatively identified with the historical comets C/1490~Y1 and C/1385~U1. In the present work we use recovery astrometry to extend the observed arc of 2003~EH1 from 10 months to about 5 years, enough to exclude the proposed direct relationship of the asteroid with both of the comets.
\end{abstract}


\maketitle

\section{Introduction}

The Near Earth Object known as 2003~EH1 was discovered in March 2003 by B.~Skiff in LONEOS survey data, and announced in MPEC 2003-E27 \citep{2003MPEC....E...27M}. As first pointed out by \cite{2004AJ....127.3018J}, its orbit has a good similarity to that of the Quadrantid meteor stream, the last major annual meteor shower to be linked with a known comet or minor planet.\\

Before this proposed linkage, many other asteroids and comets have been suggested as possible progenitors of the shower. The most convincing of these objects, proposed by \cite{1979PASJ...31..257H}, is known as C/1490~Y1. Appearing as a bright naked-eye comet in the last days of 1490, it was extensively observed for the first two months of 1491 by Chinese, Korean, and Japanese astronomers, as reported by \cite{1962VA......5..127H}. Various parabolic orbits for this object, based on the written descriptions of the comet path in the sky, have been proposed in the past, with the most recent given by \cite{1979PASJ...31..257H}. 

There have also been various attempts to integrate the orbit of the Quadrantid stream to 1491 \citep[such as by][]{1993MNRAS.264..659W}, with the integrated orbit showing a remarkable similarity with the comet. They also proposed a possible identity of C/1490~Y1 with another historical comet, C/1385~U1, and used this possible linkage to estimate an eccentricity $e\simeq0.75$ for C/1490~Y1.\\

The comparison between various orbits is meaningful only if they are referred to the same epoch, because the action of various gravitational perturbations is often significant on a time span of a few centuries. This caution is especially important in this case, because the orbit of 2003~EH1 has an aphelion around the orbit of Jupiter, therefore it often enters a region of space in which strong perturbing effects are acting. This comparison can only be done by integrating the orbit of 2003~EH1 to 1491 (or 1385), the date of the perihelion of the comets, because the two cometary orbits are both too uncertain to be integrated forward meaningfully. 

The first attempt to perform such an integration to 1491 was done by \cite{2004EM&P...95...11W,2004MNRAS.355.1171W}, using observations of 2003~EH1 spanning a time period of less than one year. They sampled the actual uncertainty region for 2003~EH1 with 3500 synthetic clones, and integrated their orbits back to the 1491 epoch. The main result of their work is that the integrated position of the clones is extremely variable, but at least one of them had characteristics closely resembling those of comet C/1490~Y1, and its path in the sky also closely matches the one observed for the comet. However, most of the integrated orbits predicted the object to be around aphelion, and not close to the perihelion at which comet C/1490~Y1 was observed. They didn't attempt any integration to 1385, to compare the orbit with C/1385~U1, but they used the eccentricity inferred from the identification in their comparisons.\\

In April 2008 we recovered 2003~EH1 using the University of Hawaii $2.2 \un{m}$ telescope on Mauna Kea, Hawaii, equipped with a Tektronix 2048x2048 CCD camera. We observed it for two nights and extracted 7 accurate astrometric positions from our set of images. These data extend the observed arc by a factor of more than $5$ and should allow a comparable reduction in the expected positional uncertainty for the orbital extrapolation to 1491. The result of this updated orbital comparison, referred to both the 1491 and 1385 comets, is presented in this work.\\

\section{Analysis}

\subsection{Procedure}

In our analysis we used a method similar to the one applied by \cite{2004EM&P...95...11W}. We first generated $1200$ stochastic clones of 2003~EH1 and integrated their orbits to January 1491. We repeated the same approach, creating $600$ other clones, and integrated them to 1385. The main advantage of this approach is that it doesn't require any linear approximation, and it is therefore suitable considering the large uncertainties associated with this situation.\\

The orbit generation procedure was done using the Monte Carlo feature of the software ``FindOrb'' written by B.~Gray\footnote{Downloadable at \url{http://www.projectpluto.com/find\_orb.htm}.}; the procedure creates the stochastic orbits by adding Gaussian noise of given amplitude to each astrometric observation, and then used this new set of simulated positions to compute a best-fit orbit. These orbits were then integrated back to the desired epoch, 1491 January 22 in our case \citep[the same mid-arc time used in][]{2004EM&P...95...11W} and 1385 November 01 (the perihelion date of C/1385~U1). We used a Gaussian noise model with $0.8\unsec$ as the standard deviation\footnote{This is probably a strong overestimation of the true astrometric error of our Mauna Kea observations, which are internally consistent to the $\sim0.2\unsec$ level.}. The integration takes into account the gravitational influence of all the major planets, the Moon and Pluto according to the long-term planetary ephemeris JPL DE-406, and of (1) Ceres, (2) Pallas and (4) Vesta.\\

\subsection{Results}

The evaluation of the results has been done in a way consistent with \cite{2004EM&P...95...11W}; we used the stochastically generated orbits to compute the coordinates of each clone in the plane of the sky and compared them with the position described in the ancient records.\\

As expected, our clones lay on the same line of variation found by \cite{2004EM&P...95...11W}, but the spread of the points along the line is $\sim 5$ times smaller, being $\sim20\undeg$ and mainly in declination. On the reference date of 1491 January 22 the region is centered around $\delta\simeq-58\undeg$, and the northernmost clone is at $\delta\simeq-46\undeg$, at about $85\undeg$ from the reported position of the comet. As already pointed out by \cite{2004EM&P...95...11W} these positions correspond to objects close to the aphelion of the cometary orbit.
The path in the sky is completely different from the cometary one, and the velocity of the clones is lower, as expected knowing that most of them are located around aphelion. None of the clones is consequently compatible with the information given by the ancient records.\\

The computation described above neglects the possibility of non-gravitational effects acting on the object. Under conservative assumptions\footnote{We assumed that all the non-gravitational acceleration acted on the mean motion, that the average gravitational acceleration was equal to its value at aphelion and that the object was continuously active from 1491 to the present.} we estimate that the northernmost clone will still need a non-gravitational acceleration $A \gtrsim 2\times 10^{-8}\un{ua}\un{d}^{-2}$ to have a perihelion date compatible with C/1490~Y1. This value is much larger than a typical cometary non-gravitational parameter, and comparable only with the ones measured for the most active comets known so far.\\

A similar work applied to the 1385 apparition gives a comparable disagreement: all the clones are spread on a line of variation $\sim100\undeg$ in length and centered at $\sim100\undeg$ from the expected position of C/1385~U1. The closest clone is still at about $65\undeg$ from the comet position.\\

The present orbital elements of 2003~EH1 (and their correlation matrix) are shown in \tab{OrbitEH1}, while the integrated ones are compared with those of the two historical comets in \tab{Orbits}. According to the nominal orbit, during the integration interval the object had close encounters with Venus (within $0.04 \un{ua}$ in 1517), Earth (within $0.10 \un{ua}$ in 1818) and Jupiter (within $0.25 \un{ua}$ in 1664), all capable of significant orbital perturbations.\\

Although our new observations seem to exclude the identification of 2003~EH1 with C/1490~Y1 and C/1385~U1, we cannot exclude the possibility that 2003~EH1 and the comets are fragments of some parent body that split long ago.\\

\begin{table*}
\begin{center}
\begin{tabular}{ccccccc}
\hline
 &$q~[\unp{ua}]$  & $e$  & $\omega~[\undeg]$  & $\Omega~[\undeg]$  & $i~[\undeg]$ & $T_0~[\unp{d}]$ \\%
\hline
 &$1.191238\pm0.000006$ & $0.618908\pm0.000002$ & $171.3619\pm0.0004$ & $282.9691\pm0.0003$ & $70.8014\pm0.0002$&$\mbox{2008-09-03.4541}\pm0.0005$\\
\hline
$q$  &$1$ & $-0.999$ & $+0.827$ & $-0.791$ & $+0.255$ & $+0.840$\\
$e$  &$-0.999$ & $1$ & $-0.814$ & $+0.795$ & $-0.272$ & $-0.812$\\
$\omega$  &$+0.827$ & $-0.814$ & $1$ & $-0.696$ & $0.131$ & $0.884$\\
$\Omega$  &$-0.791$ & $+0.795$ & $-0.696$ & $1$ & $-0.318$ & $-0.610$\\
$i$ &$+0.255$ & $-0.272$ & $+0.131$ & $-0.319$ & $1$ & $+0.039$\\
$T_0$&$+0.840$ & $-0.812$ & $+0.884$ & $-0.610$ & $+0.039$ & $1$\\
\hline
\end{tabular} 
\end{center} 
\caption{Orbital elements and correlation matrix for 2003~EH1 at the current standard epoch (JD 2454600.5) and mean ecliptic and equinox (J2000.0).}%
\label{OrbitEH1}
\end{table*}

\begin{table*}
\begin{center}
\begin{tabular}{cccccccc}
\hline
      Object &     Epoch$~[\unp{yr}]$  & $q~[\unp{ua}]$  & $e$  & $\omega~[\undeg]$  & $\Omega~[\undeg]$  & $i~[\undeg]$ & $T_0~[\unp{d}]$ \\%
\hline
C/1490~Y1 & 1491.1 & $0.761$ & $0.75$ & $164.9$ & $280.2$ & $73.4$ &$\mbox{1491-01-17.9}$\\
2003~EH1  & 1491.1 & $0.57^{+0.07}_{-0.01}$ & $0.82^{+0.01}_{-0.02}$ & $164.2^{+0.1}_{-1.4}$ & $286.2\pm0.1$ & $66.0^{+0.7}_{-1.3}$& $\mbox{1490-04-13}^{+90}_{-370}$\\
C/1385~U1 & 1385.8 & $0.79$ & $0.75$ & $182$ & $289$ & $103$ &$\mbox{1385-11-01}$\\
2003~EH1  & 1385.8 & $0.49^{+0.03}_{-0.07}$ & $0.84\pm0.01$ & $163.4^{+1.0}_{-0.7}$ & $286.8^{+0.4}_{-0.2}$ & $63.1^{+1.2}_{-1.3}$& $\mbox{1386-12-29}^{+600}_{-1400}$\\
\hline
\end{tabular} 
\end{center} 
\caption{Integrated orbital elements of the asteroid compared with the two historical comets (mean ecliptic and equinox of J2000.0). The error bars are the $99.7~\%$ limits for the distribution of each element. Most of the distributions are strongly non-Gaussian, and some of them are bimodal, presumably due to close planetary encounters.}%
\label{Orbits}
\end{table*}

\section{Conclusions}

The results obtained in this work seem to exclude the proposed identification of comet C/1490~Y1 and C/1385~U1 as historical cometary apparitions of the asteroid 2003~EH1. None of the integrated orbits matches the position and the motion of the 1491 and 1385 objects. 

The identity between the two historical comets obviously cannot be excluded by this analysis. Similarly, the possible identification of 2003~EH1 as the progenitor of the Quadrantid meteor shower is still plausible, although it now lacks of any confirmed cometary activity from the parent body.\\

\section*{Acknowledgments}
Our observations of 2003~EH1 were funded by grant AST 0709500 from the U.S. National Science Foundation.

\bibliographystyle{mn2e}
\bibliography{Articles}

\end{document}